\documentclass[amssymb,amsmath,aps,showpacs,twocolumn,floatfix,nofootinbib,showpacs,balancelastpage]{revtex4}

\usepackage{graphicx}
\usepackage{color}
\usepackage{soul}
\usepackage{latexsym}
\usepackage{epsf}
\usepackage{amsmath}

\def\beq{\begin{equation}}
\def\eeq{\end{equation}}
\def\beqn{\begin{eqnarray}}
\def\eeqn{\end{eqnarray}}
\def\bea{\begin{eqnarray}}
\def\eea{\end{eqnarray}}
\def\be{\begin{equation}}
\def\ee{\end{equation}}

\begin{document}

\voffset 1.25cm

\title{ Detecting light long-lived particle produced by cosmic ray}

\author{Peng-fei Yin $ ^1$ and Shou-hua Zhu $ ^{1,2}$ }

\affiliation{ $ ^1$ Institute of Theoretical Physics $\&$ State Key
Laboratory of Nuclear Physics and Technology, Peking University,
Beijing 100871, China \\
$ ^2$ Center for High Energy Physics, Peking University, Beijing
100871, China }

\date{\today}

\begin{abstract}

We investigate the possibility of detecting light long-lived
particle (LLP) produced by high energy cosmic ray colliding with
atmosphere. The LLP may penetrate the atmosphere and decay into a
pair of muons near/in the neutrino telescope. Such muons can be
treated as the detectable signal for neutrino telescope. This study
is motivated by recent cosmic electron/positron observations which
suggest the existence of O(TeV) dark matter and new light O(GeV)
particle. It indicates that dark sector may be complicated, and
there may exist more than one light particles, for example the dark
gauge boson $A'$ and associated dark Higgs boson $h'$. In this work,
we discuss the scenario with $A'$ heavier than $h'$ and $h'$ is
treated as LLP. Based on our numerical estimation, we find that the
large volume neutrino telescope IceCube has the capacity to observe
several tens of di-muon events per year for favorable parameters if
the decay length of LLP can be comparable with the depth of
atmosphere. The challenge here is how to suppress the muon
backgrounds induced by cosmic rays and atmospheric neutrinos.

\end{abstract}


\maketitle

\section{Introduction}

The pursuit of new light boson (LB) with mass much lighter than W
and Z bosons in the standard model(SM) has a long history. A well
known example is a light gauge boson under an extra $U(1)$ gauge
group beyond the SM gauge group $SU(3)_c\otimes SU(2)_L\otimes
U(1)_Y$. However such kind of new LB is stringently constrained
since the current precise measurements are in excellent agreement
with the SM predictions. Obviously the interaction between the new
LB and SM sector should be tiny to evade current constraints. With
the same reason detecting LB experimentally is a very challenging
task.

Recently the LB attracts more attention due to the new cosmic
observations by PAMELA \cite{Adriani:2008zq}, ATIC \cite{:2008zz}
and Fermi \cite{Abdo:2009zk}. PAMELA collaboration reported excess
in the positron fraction from 10 to about 100 GeV but absence of
anti-proton excess \cite{Adriani:2008zq}. It is still consistent
with the results released by ATIC \cite{:2008zz} or Fermi
\cite{Abdo:2009zk} experiments. The afterwards investigations
suggested that these new observations need new source of
electron/positron which may come from the dark matter (DM). The DM
particles might annihilate or decay into electron/positron in the
halo today. Moreover man began to realize the possible connection
between the heavier DM at O(TeV) and the LB at O(GeV). If the
annihilating DM produces the extra electron/positron, there is a
mismatch between the DM annihilation cross section expected in the
epoch of freeze-out and that required to account for recent new
observations. Namely the present DM annihilation cross section is
too small.  The new O(GeV) LB, via the so-called Sommerfeld
enhancement, can fill the gap. Moreover, in order to be consistent
with only electron/positron excess, LB decays preferably into
charged leptons \cite{ArkaniHamed:2008qn}.

Generally speaking such new boson may be scalar, pseudoscalar or
gauge boson. The interactions between the LB and SM sector could
arise from the mixing between the LB and photon or Higgs. It is
quite interesting to investigate how to detect such kind of LB. Many
authors have studied how to produce such LB and detect them at
colliders, namely low-energy collider, large hadron collider or
fix-target experiment. For the collider search, one often requires
the life time of LB is short, as a consequence the charged leptons
as the LB decay products could be observed at the detector. Such
charged leptons could be clearly and easily identified. The
construction of realistic model of dark sector showed that dark
sector can be more complicated. The assumption that LB is
short-lived might be incorrect. If dark sector contains an array of
LBs besides a light gauge boson, some lighter ones may be long-lived
due to the suppressed interaction with SM sector. Several recent
works provided an interesting approach to search such long-lived
particle (LLP)
\cite{Batell:2009zp,Schuster:2009au,Schuster:2009fc,Meade:2009mu}.
The DM trapped inside the Sun/Earth would annihilate into LBs. If
the LB has a long lifetime, it can travel through the Sun/Earth and
decay into gamma rays or charged leptons which could be observed.

In this paper we point out another possible way to search such kind
of LLP via the high energy cosmic rays. The high energy cosmic rays
interact with atmospheric nucleons every second and can be treated
as a natural and costless high energy hadron collider. If a proton
with energy of $E\sim 10^4$ GeV in cosmic ray collides with an
atmosphere nucleon, the center-of-mass energy $\sqrt{s}$ is
approximately $\sqrt{2m_N E}\sim 10^2 $ GeV. Such mechanism can
copiously produce LLPs through $pN$ collisions. If the lifetime of
LLP is appropriate, it can penetrate the atmosphere and arrive at
the neutrino telescope. The subsequently decay of LLP into a pair of
muons can be observed by the telescope
\cite{Batell:2009yf,Bjorken:2009mm}.

If LLP decays near the detector, the difficult task is to
distinguish the signal muon pair from the huge muon backgrounds
which are produced by high-energy cosmic rays through hadron(for
example $\pi$) decays and/or QED process. Provided that the time
information of muon is precisely recorded, one can identify two
signal muons which are supposed to arrive at detector at the same
time, not heavily polluted by two irrelevant coincident parallel
muon events. In order to suppress the muon backgrounds, the analysis
focused on quasi-horizontal events might be important
\cite{Illana:2006xg,Ahlers:2007js}. A more optimistic case is that
LLP decays inside the detector with an 'obvious' decay vertex. The
challenge here is how to distinguish almost parallel di-muon from
the single muon. Another possible case is that for a single
collision between cosmic ray particle and atmosphere nucleon, more
than one LLP are produced. In this case the multi muon pairs appear
in detector at the same time will be a significant signal.

In this work, we assume the LB has a long lifetime to penetrate
through the atmosphere, some mechanisms which satisfy this
requirement will be discussed. We consider the LB production by the
cosmic rays and simulate di-muon events from LB decay. In order to
detect such high energy muons, we focus on the large volume neutrino
telescope IceCube \cite{Klein:2008px} which will reach an effective
detecting area in square kilometers. The detector are installed
under the ice surface in a depth of 1.4 km in order to suppress muon
backgrounds produced in atmosphere. Moreover, there is a extension
of IceCube namely DeepCore which is still under construction
\cite{Resconi:2008fe}. DeepCore will be installed in the more deeper
location which will suffer less above-mentioned muon backgrounds.

This paper is organized as following. In the  section II, we
describe two classes of models which contain LLPs, discuss the main
LLP production processes and calculate the LLP production cross
section in $pN$ collision. We utilize PYTHIA to do the calculation
and simulate the LLP events. In Section II, we also investigate the
LLP production flux produced by primary cosmic ray. We find that for
a LLP production process with O$(10)$ pb, the production rate of LLP
can reach to O$(10^4)$ per square kilometer and per year. In Section
III, we investigate the possibility to detect di-muon signal from
LLP decay in the neutrino detector. The conclusions and discussions
are given in the last section.

\section{The production of LLP}

\subsection{The model}

In the models discussed in Ref.\cite{Batell:2009zp,Schuster:2009au},
the dark sector includes both weakly-interacting-massive-particle
(WIMP) and LB under a certain new gauge group. For the simplest case
with an extra U(1) group, the dark sector can interact with SM
sector though kinetic mixing, namely a new $U(1)$ gauge field
$A'_\mu$ could mix with $U(1)_Y$ field $A_\mu$ in the SM. In
addition there is another possible mixing between the SM Higgs field
$H$ and scalar $h'$ which will break extra U(1) group to induce mass
to $A'$. The Lagrangian can be written as,
\begin{eqnarray}
\mathcal{L} &=&  - \frac{1}{4}F_{\mu \nu}^{\prime \; 2}
+ \frac{\kappa }{2}F_{\mu \nu}^{\prime} F^{\mu \nu}+|D_\mu h'|^2 - V(h') \nonumber\\
&+&   \lambda _{h'H} (h'^\dag  h')(H^\dag  H)+
\mathcal{L}_{DM}+\mathcal{L}_{SM}.
\end{eqnarray}
Here $\kappa $ and  $\lambda _{h'H}$ are the two mixing parameters
which will be determined (constrained) by experiments.
$\mathcal{L}_{DM}$ is the lagrangian of DM which includes the
O$(TeV)$ DM kinetic and mass terms and its gauge interaction.

After the spontaneous breaking of extra U(1) group, the $A'$ and
$h'$ will get the mass of $m_{A'}$ and $m_{h'}$. If scalar mixing
$\lambda _{h'H}$ is neglected, the lifetime of LB is determined by
$m_{LB}$ and gauge kinetic mixing parameter $\kappa$. For light
gauge boson mainly decay into charged leptons, the travel distance
can be approximately estimated as $l=\gamma c\tau \sim \gamma
c/(\alpha \kappa^2 m_{A'}) \sim 10^{-5}$ m
$(\gamma/10^3)(\kappa/10^{-3})^{-2}(m_{A'}/1GeV)^{-1}$. Typically
the light gauge boson is not a long-lived particle. However for the
dark Higgs boson and provided that $m_A'>m_h'$, dark higgs will
decay into SM fermions through the triangle diagrams. The decay
width is suppressed by a factor of $\kappa^{-4}$
\cite{Batell:2009yf}. Therefore the dark higgs can be a typical
long-lived particle (LLP) with the decay length as large as $10^{7}$
km \cite{Batell:2009zp}. In such kind of models, the main LLP
production mode is scalar-strahlung process $pp \rightarrow
A'^{*}+X\rightarrow A'h'+X$ ( $X$ denotes anything).

The dark sector could be more complicated. As discussed in Ref.
\cite{Baumgart:2009tn,Zhang:2009dd}, if the dark sector has a more
complex gauge group configuration, there might exit a series of LBs
including the LLP. In such case, the high energy $pN$ collisions may
copiously produce the extra gauge boson $A'$ and the production rate
depends on the mixing parameter $\kappa $ and  $m_{A'}$. $A'$
subsequently decays quickly into LLP, namely the dark Higgs boson
$h'$. The process can be depicted as $pp \rightarrow A'+ X
\rightarrow h'a'+X$ ($a'$ represents another light gauge boson or
light pseudo scalar which we do not discuss further its feature
here. For simplicity we take the mass of $h'$ and $a'$ to be equal).
If LLP propagates some distance which depends on its lifetime and
decays into charged leptons as motivated by cosmic electron/positron
data, we can utilize neutrino telescope to observe such kind of
muons. Note that the LLP interacts with usual matter weakly, it can
penetrate into the Earth without loss of energy.

In this case, we do the calculation as model-independent as possible
and the input parameters are chosen as kinetic mixing $\kappa$ and
the mass of extra gauge boson $m_{A'}$ which determine $A'$
production rate, mass $m_{h'}$ and lifetime $\tau$ of $h'$.
Typically branching ratio of $A'$ into dark sector is much bigger
than that into SM sector. Thus in our numerical simulation we assume
that $A'$ decays only into $h'$. We also assume the branching ratio
of $h'$ into muon pair is 1. In fact, such branching ratios are
calculable in the model we mentioned above, the detailed
calculations can be found in the Ref.
\cite{Schuster:2009au,Batell:2009yf}. Here we just treat them as
free parameters, and our results could be adjusted to satisfy
specific model by multiplying a factor of $Br(A'\rightarrow
h')Br(h'\rightarrow \mu^+\mu^-)$.

\subsection{ Simulation of LLP production}

\begin{figure}[h]
\centering
\includegraphics[totalheight=2.5in]{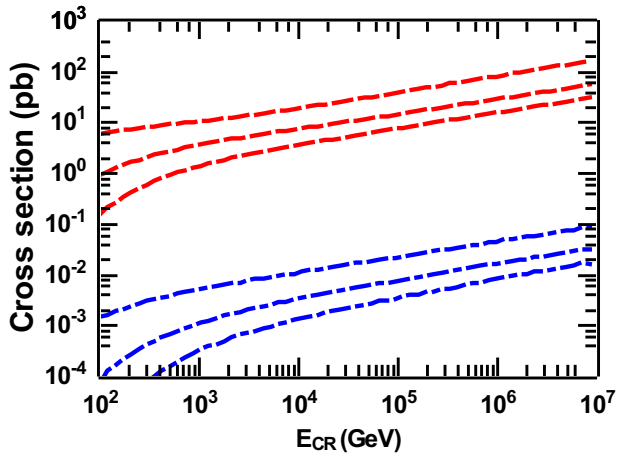}
\caption{Cross section of LLP production in $pp$ collision as a
function of cosmic primary proton energy. The dash lines and the
dashed-dot lines represent process $q\bar{q}\rightarrow A'$ and
$q\bar{q}\rightarrow h'A'$ respectively. For the lines with same
shape, the upper and lower one represent three benchmark point
$(m_{A'},m_{h'})=$ (1.2 GeV, 0.4 GeV), (2.1 GeV, 0.7 GeV), (3.0 GeV,
1.0 GeV) respectively. } \label{cspp}
\end{figure}

We utilize PYTHIA \cite{Sjostrand:2006za} to simulate the LLP
production events and calculate the cross section. The main free
parameters are the mass of $A'$ and $h'$, kinetic mixing parameter
$\kappa$, and dark sector gauge coupling $\alpha'$ for the
Higgs-stahlung process. Here we choose three benchmark points with
$(m_{A'},m_{h'})=$ (1.2 GeV, 0.4 GeV), (2.1 GeV, 0.7 GeV), (3.0 GeV,
1.0 GeV) and $\alpha'=\alpha$ for simplicity. The $\kappa$ is
severely constrained by experiments, and we choose $\kappa$ as
$10^{-3}$ which is still allowed \cite{Bjorken:2009mm}. In addition,
for the case $m_{h'}<m_{A'}$ we discussed here, the $\kappa$ could
not be very small to avoid destroying the success of BBN. The
lifetime of $h'$ must be shorter than 1 second, and the $\kappa$
should not be very smaller than $10^{-4}$ \cite{Batell:2009yf}.

For the scalar-strahlung process, the cross section at parton level
is given by (neglecting the mass of parton),
\begin{equation}
\hat{\sigma}(\hat{s})|_{q\bar{q}\rightarrow A'h'}=\frac{8\pi
Q_q^2\alpha\alpha'\kappa^2}{9}
\frac{k}{\sqrt{\hat{s}}}\frac{k^2+3m_{A'}^2}{(\hat{s}-m_{A'}^2)^2},
\end{equation}
where $\hat{s}=x_1 x_2 s$ is the center of mass energy of partons
with momentum fractions of $x_1$ and $x_2$,
$k=\sqrt{(\hat{s}-m_{A'}^2-m_{h'}^2)^2-4m_{A'}^2m_{h'}^2}/(2\sqrt{\hat{s}})$
is the momentom of $A'$ in the center of mass frame. The total cross
section is
\begin{equation}
\sigma_{pN}=\int dV \sum_{q}f^q_p(x_1)
f^{\bar{q}}_N(x_2)\hat{\sigma}(\hat{s})_{q\bar{q}},
\end{equation}
where $d V$ represents $dx_1 dx_2$, $q$ denotes sum over all the
quark and anti-quark and $f$ is the parton distribution function
(PDF). We use the CTEQ6M PDF \cite{Pumplin:2002vw} here and set the
factorization scale $Q^2=\hat{s}$. To do a Monte Carlo calculation
\cite{barger}, the variables $x_1$, $x_2$ are randomly chosen within
the ranges $(m_{A'}+m_{h'})^2/s \leq x_1 \leq 1$,
$(m_{A'}+m_{h'})^2/(x_1s) \leq x_2 \leq 1$. In order to improve the
convergence, we technically define new integration variables as $dV'
= d \ln x_1 d \ln x_2$. The integration is then transformed into
\begin{equation}
\sigma_{pN}=\frac{1}{N_{tot}} \sum_i \left[\sum_{q}x_1 f^q_p(x_1)
x_2f^{\bar{q}}_N(x_2)\hat{\sigma}(\hat{s})_{q\bar{q}} dV'\right],
\end{equation}
where $i$ denotes i-th configuration of cross section.

For the single $A'$ production which is a $2\rightarrow 1$ process,
the cross section contains one $\delta$ function which fixes
$\hat{s}=m_{A'}^2$. We transform integration variables $x_1$, $x_2$
to $\hat{s}$, $y$ through $dx_1dx_2=d\hat{s}dy/s$ and integrate out
the $\delta$ function. The total cross section can be written as
\begin{equation}
\sigma|_{pN\rightarrow A'X}=\int \sum_{q} \frac{4\pi^2
Q_q^2\alpha\kappa^2}{3m_{A'}^2} x_1f^q_p(x_1) x_2f^{\bar{q}}_N(x_2)
dy
\end{equation}
where $x_1=m_{A'}e^y/\sqrt{s}$ and $x_2=m_{A'}e^{-y}/\sqrt{s}$. From
$x_{1,2}\leq 1$, we choose $y$ in the range of $-\ln
(\sqrt{s}/m_{A'})\leq y \leq \ln(\sqrt{s}/m_{A'})$.

We show the LLP production cross sections for the $pp$ collision in
Fig. \ref{cspp}. From the figures we can see that the
scalar-strahlung process cross sections are much smaller than those
of $A'$ resonance process. It is simply because the scalar-strahlung
process is the $2 \rightarrow 2$ process which has more power of
coupling and smaller phase space. Numerically in the energy region
of O$(100)$ GeV, the single $A'$ production is several orders of
magnitude larger than that the scalar-strahlung process. From the
Fig. \ref{cspp} we can also conclude that cross section is very
sensitive to the mass of $A'$. This is quite understandable provided
the quick rise of $q\bar q$ luminosity for the lower mass $A'$. Note
that flux of cosmic ray decreases quickly with the increment of its
energy, thus the behavior for low energy cosmic ray plays the major
role to produce LLP.

In our simulations, we randomly generate muon pairs in the $h'$ rest
frame, then boost them to the $A'$ rest frame and lab frame.

\subsection{Flux of long-lived particle produced by high energy cosmic rays}

\begin{figure}[h]
\centering
\includegraphics[totalheight=2.2in]{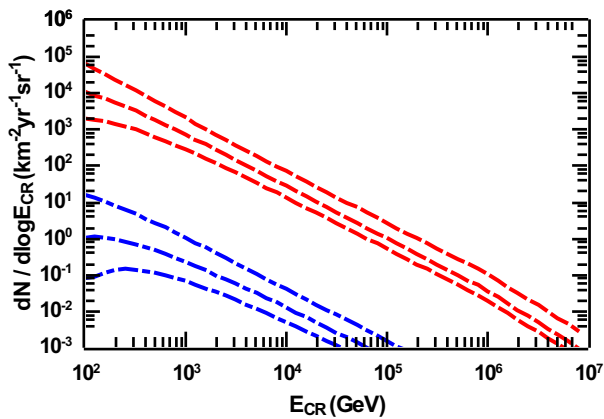}
\caption{The LLP differential production rate as a function of the
energy of primary cosmic ray. The notations are same as Fig.
\ref{cspp}.} \label{num}
\end{figure}

In this subsection, we evaluate the LLP production rate by high
energy cosmic ray. From several GeV to energy range above $10^6 $
GeV, the flux of primary nucleons in the cosmic rays is
approximately written as \cite{Amsler:2008zzb}
\begin{equation}
\Phi_{N}(E)\approx 1.8(E/GeV)^{-\alpha} \frac{nucleons}{cm^2 \; s \;
sr \; GeV},
\end{equation}
where the differential spectral index $\alpha$ is $2.7$ under $10^6
$ GeV,  $3.0$ from $10^6 $ GeV to $10^{10}$ GeV, and finally $2.7$
above $10^{10}$ GeV. The main component of primary cosmic nucleons
is proton. To produce energetic LLP, the energy of cosmic rays is
required to be as high as or above $10^2 $ GeV. For secondary
hadrons induced by such energetic cosmic ray, the interaction length
may be smaller than the decay length in the atmosphere. Thus the
collisions between the atmosphere nucleons and secondary mesons
induced by primary cosmic hadrons are important \footnote{For a
simple and conservative estimation, this work does not include the
effects of secondary hadrons. The influence of secondary hadrons to
final LB flux can be included by multiplying an extra O(1) factor.
This factor is determined by the flux of secondary hadrons
\cite{Illana:2006xg}.}. The flux of new LLP particles can be
estimated by \cite{Illana:2006xg}
\begin{equation}
\Phi_{h'}=\sum_h \int^{E_{max}}_{E_{min}}dE \; \Phi_{h}(E)\; {\cal
P}^h_{h'}(E),
\end{equation}
where $h$ denotes the primary cosmic hadrons and secondary nucleons,
pion and kion, $\phi_h$ is the flux of hadron, and ${\cal
P}^h_{h'}(E)$ is the probability of producing LLP $h'$ in one
collision. ${\cal P}^h_{h'}(E)$ is approximated as ${\cal
P}^h_{h'}(E)\approx A \sigma^{hN}_{h'}/\sigma_T$
\cite{Illana:2006xg}, where $A\sim 14.6$ is the average nucleon
number of a nuclei in air, $\sigma^{hN}_{h'}$ is the cross section
to produce $h'$, and $\sigma_T$ is the total cross section of cosmic
hadrons in atmosphere about O$(10^2)$ mb which is approximately
parameterized as $\sigma_T \approx C^h_0+ C^h_1 \ln(E/GeV)+ C^h_2
\ln^2(E/GeV)$ \footnote{In fact, the total cross sections of high
energy cosmic hadrons are not precisely determined. There are
several models which are consistent with the exiting observations
roughly, but more data are needed to give a more concrete result.
The detailed discussions about different models and parameter
selections can be found in Ref. \cite{Heck:1998vt}. Here we use the
result of VENUS model \cite{Werner:1993uh} given in Ref.
\cite{Heck:1998vt} as an acceptable approximation from $10^2$ GeV to
$10^7$ GeV. For present experiment results and comparisons between
different theoretical predictions, one can see the Ref.
\cite{Aielli:2009ca} and references therein.}. Requiring final muons
from LLP decay with large energy and flux, we choose the cosmic ray
energy region from $10^2$ GeV to $10^7$ GeV.

We can firstly estimate the LLP production rate in order of
magnitude as
\begin{equation}
\phi_{h'}\sim10^3(\frac{\sigma^{hN}_{h'}}{1pb})
(\frac{\sigma_T}{300mb})^{-1}(\frac{E_{min}}{100GeV})^{-1.7}km^{-2}yr^{-1}sr^{-1}.
\end{equation}
The formula indicates that for the single $A'$ production cross
section of $10$ pb, the cosmic ray will produce at least $10^4$ LLPs
per year and per square kilometer. In Fig. \ref{num} we show the
results of LLP differential rate as a function of primary cosmic ray
energy. From the figure we can see that the event rate can reach
O$(10^2)$ to O$(10^4)$ $km^{-2}yr^{-1}sr^{-1}$. However for the
scalar-strahlung process, the event rate is very low from
O$(10^{-1})$ to O$(10)$ due to the small cross section and we do not
discuss this process below.

\section{Detecting long-lived particle at high energy neutrino detector}

In this section we will discuss the possibility of detecting such
LLP at neutrino detector. If the LLP penetrates through atmosphere
and decays into muons near or inside detector, this signal can be
observable at neutrino telescope. In the last section, we have
evaluated the $h'$ production rate. For simplicity, we assume that
all $h's$ are produced at the upper atmosphere. Moreover, the $h'$
flux can be treated as isotropic due to the isotropy of cosmic ray.
The long-lived $h'$ interacts with SM particle weakly, thus it is
safe to neglect the energy loss before it decays into muon pair.

The lifetime of $h'$ is model-dependent, and it is treated as a free
parameter here. The probability of a particle decays between two
point is given by \cite{Meade:2009mu}:
\begin{equation}
P_{decay}=e^{-D/l}-e^{-(D+d)/l}=e^{-D/l}(1-e^{-d/l}), \label{pdecay}
\end{equation}
where $D$ is the distance between the production point and entry
point, $d$ is the distance between two point, $l=\gamma v \tau$ is
the decay length of particle. Here $D$ is approximately $10\sim20 $
km, and it is the depth of atmosphere
 plus distance from detector to horizon; $d$ is the neutrino telescope's size
 (if the detector
has the capacity to recognize LLP decay events nearby, $d$ can be
larger). If $D$ is far larger than $d$ and $l$, the decay
probability is roughly $e^{-D/l}$; if the decay length $l$ is far
larger than $D$ and $d$, the decay probability is $d/l$. Therefore,
it is difficult to observe the LLP decay if the decay length of LLP
is too large or too small. The most promising case is that the decay
length $l$ is comparable with $D$.

As mentioned in the introduction if LLP decays outside the neutrino
detector, the atmosphere muon background is huge. These muons are
generated mainly from secondary charged pion and kaon decay. Most
muons are in lower energy regime due to the relatively large cosmic
ray flux in this energy region and energy lost in matter. The
formula of atmosphere muon flux is similar to that of cosmic ray
which is approximately given by \cite{Amsler:2008zzb}
\begin{equation}
\phi_{\mu}(E_{\mu})\approx \frac{0.14 E_{\mu}^{-2.7}}{cm^2 \;s
\;sr\; GeV}( \frac{1}{1+\frac{1.1
E_{\mu}\cos\theta}{115GeV}}+\frac{0.054}{1+\frac{1.1
E_{\mu}\cos\theta}{850GeV}}), \label{muon}
\end{equation}
where $\theta$ is zenith angle (here $\theta \leq 70^\circ $), and
this formula is valid when the probability of the muon decay can be
neglected (i.e. $E_\mu>100/\cos \theta$ GeV). From the
Eq.(\ref{muon}), we can see that in the energy region O$(10^2)$ GeV
the atmospheric muon flux is only one order of magnitude smaller
than primary cosmic ray, while the LLP flux is about ten orders
smaller. Requiring that two muons arrive at detector in a tiny time
window could effectively reduce background which contains two
uncorrelated muons. Note that lots of SM processes can produce muon
pair events directly \cite{Illana:2009qv}. For example, one single
shower contains many hadrons, and muons from two hadrons decay may
be treated as a pair of muons. In addition the electro-weak
Drell-Yan process can also produce muon pair directly. In a word, if
the LLP does not decay inside the detector, it is very challenging
to distinguish signal from the atmospheric di-muon backgrounds.

However there is still hope to detect muon pair from LLP decay near
the detector. The energy of LLP is typically less than $1$ TeV. In
such energy region, most of the atmospheric muons are absorbed in
the solid/liquid matter. For example, for $10$, $10^2$ and $10^3 $
GeV muons, the muon range is $0.05$, $0.41$ and $2.45$ km.w.e
respectively \cite{Groom:2001kq}. The neutrino detector is often
installed in deep underground with shield of several km.w.e.
rock/water/ice. Such shield can prohibit the low-energy atmospheric
muons to arrive at the detector, especially for direction with large
zenith angle and even quasi-horizontal direction. Detecting LLP
decay in these directions is more promising. Thus if the LLP decays
not far from detector, the resulting di-muon may be identified.

To detect the clean LLP signal, we expect that LLP happens to decay
inside the detector and we can observe a pair of "suddenly" appeared
muons. In order to recognize di-muon event, we require that the
energy of muons must be above the detector's threshold energy.
Moreover, because the high energy atmospheric neutrino may produce
single high energy muon when traveling through the detector, we also
require that the two tracks of di-muon should be identified
separately \footnote{The authors of Ref. \cite{Meade:2009mu} have
pointed out that it is possible to distinguish di-muon signal from
single muon even the separation of di-muon is not large enough. They
provided two handles to identify di-muon signal with energy above
critical energy utilizing the different characteristics of Cherenkov
radiation.}. In our simulations, we require that the angle between
two muon tracks is greater than O$(10^{-4})$.

It needs to mention another background arising from atmospheric
neutrino, which can induce di-muon events inside the detector
through inelastic $\nu N$ collision. $\nu N$ collision can produce
muon plus charm hadron and the charm hadron will subsequently induce
another muon via semi-leptonic decay \cite{Albuquerque:2006am}. Such
two muons will be identified as a di-muon event. Such di-muon event
rate may be larger than the LLP decay due to large atmospheric
neutrino flux (for example, the atmospheric muon neutrino flux
around $100$ GeV is about O$(10^{-4})m^{-2}s^{-1}sr^{-1}GeV^{-1}$).
However, associated hadronic shower in $\nu N$ collision could be
utilized to suppress such backgrounds.

\begin{figure}[h]
\centering
\includegraphics[totalheight=2.4in]{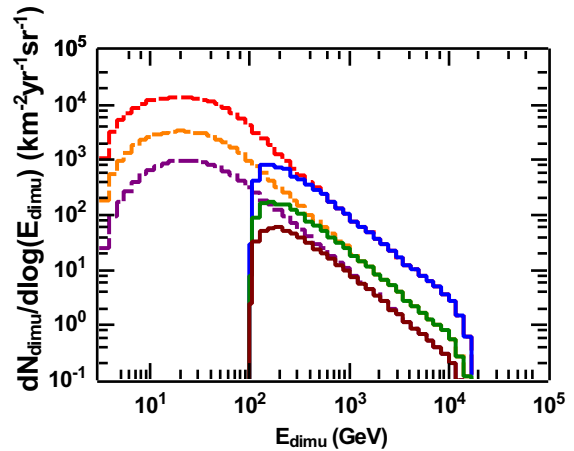}
\caption{Di-muon event rate as a function of di-muon energy. The
dash lines and the solid lines correspond to different energy
threshold of $3.2$ GeV for Super-Kamiokande and $100$ GeV for
IceCube respectively. For the lines with same shape, the upper and
lower one represent three benchmark point $(m_{A'},m_{h'})= $(1.2
GeV, 0.4 GeV), (2.1 GeV, 0.7 GeV), (3.0 GeV, 1.0 GeV) respectively.
\label{thmu}}
\end{figure}

\begin{figure}[h]
\centering
\includegraphics[totalheight=2.2in]{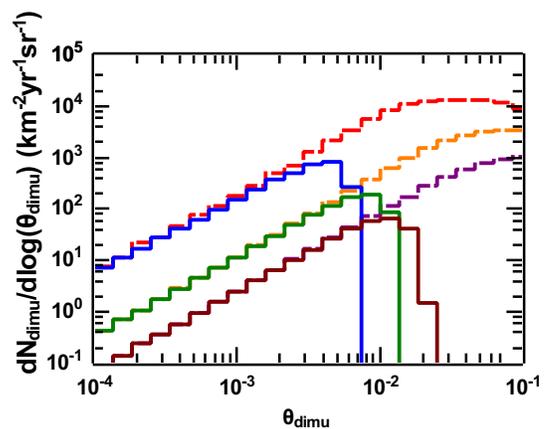}
\caption{Di-muon event rate as a function of separation angle
between two muons. The notations are same as Fig. \ref{thmu}.
\label{emu} }
\end{figure}

Now we scrutinize the possibility of detecting LLP at the large
volume Cherenkov detector IceCube. IceCube is a $1 km^3$ detector
which consists of 4800 digital optical modules and is installed
between depth of $1450$ m and $2450$ m in the south pole
\cite{Klein:2008px}. In Fig. \ref{thmu} and Fig. \ref{emu}, we show
the di-muons rate as the functions of the di-muon energy and the
separation angle. Here we assume all the LLPs decay in the detector.
We take the each muon energy threshold of IceCube to be $50$ GeV and
the separated angle of di-muon greater than $10^{-4}$. For
comparison, we also show the rate of di-muon events with energy
above $3.2$ GeV which corresponds to the threshold at the
Super-Kamiokande detector \cite{Ashie:2005ik}. From Fig. \ref{thmu}
we find that di-muon event rate with energy larger than $100$ GeV
could be O$(10^3)$ per square kilometer, per year. From Fig.
\ref{emu} we find that the separated angle of most di-muons is about
O$(10^{-2})$. For the Super-Kamiokande detector with $R=16.9$ m,
$H=36.2$ m and energy threshold $1.6$ GeV, it is suitable for
detecting the low energy di-muon events with large flux. However it
is difficult to distinguish the di-muon events from single muons due
to the limited volume.

\begin{figure}[h]
\centering
\includegraphics[totalheight=2.3in]{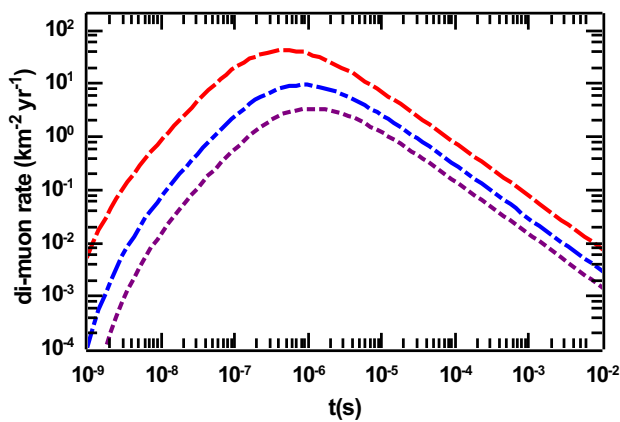}
\caption{Di-muon rate as a function of the lifetime of LLP, where
$(m_{A'},m_{h'})=$ (1.2 GeV, 0.4 GeV), (2.1 GeV, 0.7 GeV), (3.0 GeV,
1.0 GeV) from top to bottom. The size of detector are taken as $1$
km and the zenith angle is less than $\theta\leq70^\circ$. We choose
each muon energy threshold of detector to be $50$ GeV and the
separated angle of di-muon to be greater than $10^{-4}$.}
\label{finnum}
\end{figure}

In Fig.\ref{thmu} and Fig. \ref{emu}, we did not include effects of
LLP decay. The Fig. \ref{finnum} shows the di-muon rate as a
function of the lifetime of LLP. We assume that the LLP is randomly
produced between the altitude of $15$ km and $20$ km and the LLP
decay length is taken as $l = c\tau E/m_{h'}$. The size of IceCube
is chosen as $1$ km. The effective area of detector is not taken
into account since it needs the detailed simulation including
realistic experimental trigger and condition.

The detector is insensitive to the direction for low zenith angle.
The reason is that for a given lifetime $\tau$, the LLP has similar
$l$ and $P_{decay}$ in these directions. Therefore we show the rate
in the directions with zenith angle below $70^\circ$. We choose each
muon energy threshold of detector to be 50 GeV and the separated
angle of di-muon to be greater than $10^{-4}$ here, and find that
the di-muon rate can reach several tens for the most optimistic
cases. For example, the event rate can be 42,  9,  3 per year for
three benchmark points $(m_{A'}, m_{h'})=$ (1.2 GeV, 0.4 GeV), (2.1
GeV, 0.7 GeV), (3.0 GeV, 1.0 GeV) if the lifetime of LLP is $\sim
5\times10^{-7}s$. As mentioned above, the decay probability is
highest at $l\sim D$, thus the LLP with lifetime around $\tau \sim
10^{-6} (l/20km)(m_{h'}/1GeV)(E/10^2GeV)^{-1}$ s will induce more
di-muon events in detector.

\section{Discussions and conclusions }

In this paper, we investigated the possibility of searching
long-lived particle (LLP) produced by high energy cosmic ray
colliding with atmosphere. The LLP may penetrate the atmosphere and
decay into a pair of muons near/in the neutrino telescope. Such
muons can be treated as the detectable signal. This study is
motivated by recent cosmic electron/positron observations by
PAMELA/ATIC/Fermi. The new data suggests new source of
electron/positron which may come from O(TeV) dark matter. In order
to understand the dark matter thermal history, new light O(GeV)
particles have been proposed. It is quite natural to conjecture that
dark sector is complicated. There are more than one light particles
in dark sector, for example the dark gauge boson $A'$ and associated
dark Higgs boson $h'$. In this paper, we studied the scenario with
$A'$ heavier than $h'$ and $h'$ is treated as LLP.

We have studied the LLP production processes and found that the
promising process is single $A'$ production $q\bar{q}\rightarrow
A'$, and $A'$ subsequently decays into $h'$ rather than into SM
particles. Our numerical calculations show that for $A'$ with mass
of $1 GeV$, the production rate can reach $10^{4} km^{-2} s^{-1}
sr^{-1}$ and the final di-muon rate from $h'$ is serval tens for
some favorable parameter region. We have assumed the $\kappa \sim
10^{-3}$, $\alpha'=\alpha$ and the branching ratio of $A'\rightarrow
h' \rightarrow \mu^{+}\mu^{-}$ is $1$. For different parameter
selections, our results need to be multiplied by a factor of
$(\kappa/10^{-3})^2 Br_{A'\rightarrow h'}Br_{h'\rightarrow
\mu^{+}\mu^{-}}$. Here it is worth remarking that a complete
analysis for LLP production needs QCD correction to production
process and simulations of cosmic ray shower included secondary
hadrons and nucleons. The LLP could also be produced from secondary
meson decay directly if allowed by kinematics \cite{Batell:2009di}.
These elements will increase LLP production rate efficiently.

We simulated the signal di-muon events and calculated the rate as
functions of di-muon energy and the separation angle between two
muons. Our numerical results showed that the large volume neutrino
detector IceCube is suitable to detect LLP with lifetime about
$10^{-8}\sim 10^{-4}$ s. It is worth to mention that such parameter
region could be compatible with the constraints from fixed-target
experiments. For example, the Ref. \cite{Schuster:2009au} has
reported the constraints on LLP decay length $c \tau$ between $1$ cm
and $10^8$ cm by CHARM experiment result \cite{Bergsma:1985qz}.

The scenario proposed in this paper could be cross-checked (tested)
at the low energy $e^{+}e^{-}$ collider and/or large hadron
collider. At the collider, the LLP will escape from the detector and
act as the missing energy. For example, the well promising process
is the $\gamma A'$ associated production \cite{Yin:2009mc}. The
$\gamma+ {E\!\!\!\! /\,\,}_T$ signal can be isolated from SM
irreducible background $\gamma Z \rightarrow \gamma \nu \bar{\nu}$.
If the LLP associated production with other light bosons which decay
into charged leptons, the multi-
$e^{+}e^{-}/\mu^{+}\mu^{-}+{E\!\!\!\! /\,\,}_T$ is a cleaner signal.
Moreover, the interaction between SM and dark sector may be induced
via the mixing of Higgs fields. It implies that SM Higgs may decay
into LLP. Such possible invisible decay modes can even change our
search strategies of SM Higgs boson.

\section{ Acknowledgements}

We thank Jia Liu for helpful discussions. This work was supported in
part by the Natural Sciences Foundation of China (Nos. 10775001,
10635030).

\end{document}